\documentclass[aps,twocolumn,showpacs,floats,superscriptaddress]{revtex4-1}
\usepackage{graphicx}
\usepackage{amssymb,amsmath}
\usepackage{color}
\usepackage{bm}
\begin{document}

\author{Nikolay V. Prokof'ev}
\affiliation{Department of Physics, University of Massachusetts,
Amherst, MA 01003, USA}
\affiliation{Russian Research Center ``Kurchatov Institute'',
123182 Moscow, Russia}

\author{Boris V. Svistunov}
\affiliation{Department of Physics, University of Massachusetts,
Amherst, MA 01003, USA}
\affiliation{Russian Research Center ``Kurchatov Institute'',
123182 Moscow, Russia}


\title{Spectral Analysis by the Method of Consistent Constraints}


\date{\today}
\begin{abstract}
Two major challenges of numeric analytic continuation---restoring the spectral density, $s(\omega)$, from corresponding Matsubara correlator, $g(\tau)$---are (i)
producing the most smooth/featureless answer for $s(\omega)$, without compromising the error bars on
$g(\tau)$, and (ii) quantifying possible deviations of
the produced result from the actual answer. We introduce the method of consistent constraints that solves both problems.
\end{abstract}


\maketitle
A dynamic linear-response function can be obtained from the associated
Matsubara correlator by analytic continuation of the latter. A prototypical example is
the ground-state single-particle Matsubara Green's function in imaginary-time representation, $g(\tau)$.
If  $g(\tau)$ is specified numerically \cite{remark1}, the procedure of analytic continuation---often referred to as  spectral analysis---amounts to finding
an appropriate spectral function $s(\omega)$ related to $g(\tau)$ by an integral equation.
In our prototypical case, the equation reads (here the function $s(\omega)$ is known
to be  identically zero at $\omega < 0$ and non-negative at $\omega\ge 0$)
\begin{equation}
g(\tau)\, =\, \int_0^{\infty} {\rm e}^{-\omega \tau}s(\omega) \, d\omega \, .
\label{g_tau}
\end{equation}
The notorious difficulty of the problem comes from the fact that finding $s(\omega)$ does not reduce
to the requirement that the integral in the right-hand side of (\ref{g_tau}) reproduces
the values of $g(\tau)$ within their error bars. Being ill posed, i.e. subject to the sawtooth instability,
the problem of numerically finding $s(\omega)$  features infinitely many solutions,
ranging continuously from very smooth to extremely noisy ones. It is crucial, thus,
not only to satisfy Eq.~(\ref{g_tau}) for a specified set of $\tau$-points, but also to
guarantee that $s(\omega)$ is free of sawtooth artifacts. The following two approaches
work rather well in achieving this goal: (i) the  stochastic-optimization method (SOM) \cite{andrey}
and (ii) the maximum entropy method (MEM) \cite{MaxEnt1,MaxEnt2}.
Within SOM, one employs a stochastic process of minimizing the standard $\chi^2$-measure
to produces a large number of noisy solutions, and then takes an average of all the solutions,
which is a legitimate procedure thanks to linearity of Eq.~(\ref{g_tau}). In the statistical limit,
the outcome of SOM is a rather smooth function as the sawtooth artifacts are averaged out.
By construction, the accuracy of reproducing $g(\tau)$ with $s(\omega)$ is not compromised,
but, speaking generally, with SOM one cannot guarantee that the final result for $s(\omega)$
is the smoothest of all the functions consistent within the error bars of $g(\tau)$.
MEM is complimentary to SOM in the sense that, on one hand, it does guarantee that
the outcome for $s(\omega)$ is the best one within a certain class of smooth functions
(this class is selected by formulating a ``target" or ``default" model for the solution,
see \cite{MaxEnt1,MaxEnt2}). On the other hand, smooth solutions are produced at the expense
of a systematic bias introduced by the default model; the bias becomes more pronounced
as the error bars on $g(\tau)$ are decreased.

Existing methods of spectral analysis, with MEM and SOM as characteristic examples, seem to follow
the general principle (cf., e.g., Tikhonov-Phillips regularization methods  \cite{Tikhonov, Phillips,BSS}) that there is always a compromise between the requirement of $s(\omega)$
being as smooth as possible and the requirement of reproducing $g(\tau)$ within the error bars.

In this Letter, we show that the ``compromise principle" is a mere prejudice.
It is possible, and relatively easy (!), to meet the condition of smoothest possible $s(\omega)$
while perfectly respecting the error bars on $g(\tau)$. The  price that one has to pay for this
luxury is the necessity to introduce a feedback loop locally adjusting the smoothness constraints
on $s(\omega)$ to ensure consistency with the error bars on $g(\tau)$. More importantly, the method
of consistent constraints (MCC) has a simple built-in tool of quantifying the accuracy of  $s(\omega)$.

The issue of quantifying the accuracy of $s(\omega)$ is yet another challenge for the problem of spectral analysis.
The sawtooth instability implies that any error bar on $s(\omega)$ should necessarily be of a
{\it conditional} character. The condition which we adopt (finding it natural) is as follows.
Any function $s(\omega)$ that we include into the class of legitimate deviations from the optimal
(i.e. most smooth) solution  is subject to the constraint of not having extra qualitative
features like, e.g., extra bumps or minima (more specifically,
the constraint is to keep the same structure of sign-domains of the second derivative, see below).
With this constraint---clearly justified in the asymptotic regime of appropriately small error bars
on the function $g(\tau)$---we employ MCC to deliberately distort the function $s(\omega)$ at a certain
$\omega=\omega_*$ and see the extent to which the distortion remains consistent with the error bars of $g(\tau)$.

\begin{figure}[t]
\includegraphics[width=0.95\columnwidth]{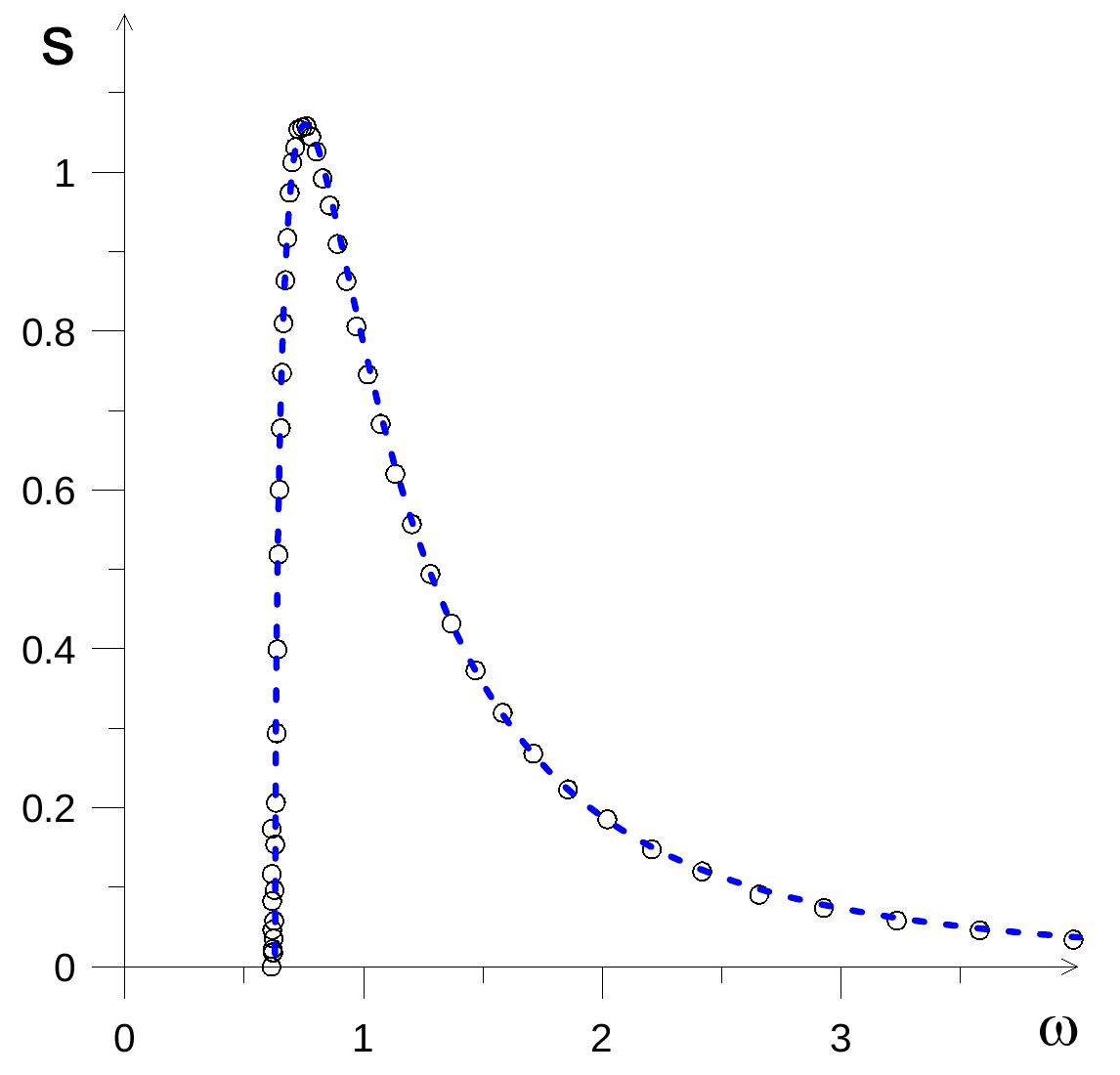}
\vspace*{-0.2cm}
\caption{\label{fig:1} Case A: The  function $s(\omega)$ (circles), obtained by spectral analysis, is very close to the exact spectral density $s_e(\omega)$ (dashed line).
The task for the error analysis is to confirm small error bars on $s(\omega)$ without knowing $s_e(\omega)$.
}
\end{figure}

\begin{figure}[t]
\includegraphics[width=0.95\columnwidth]{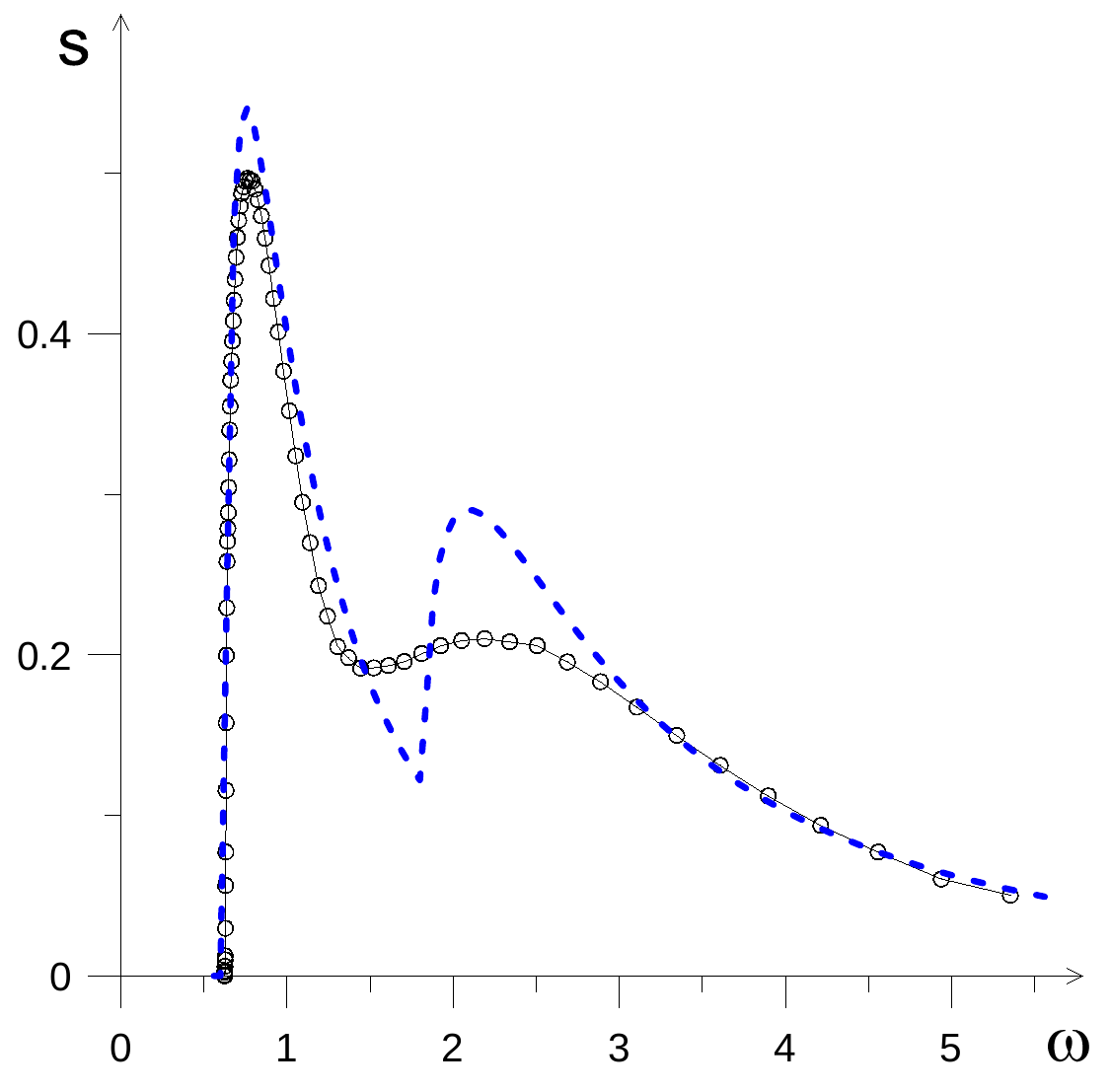}
\vspace*{-0.2cm}
\caption{\label{fig:2} Case B: Close to the second peak, the function $s(\omega)$ (circles), obtained by spectral analysis,  is substantially smother than  its exact counterpart $s_e(\omega)$ (dashed line).
The challenge for the error analysis here is to characterize possible deviations of $s(\omega)$ from $s_e(\omega)$, without knowing the latter.
}
\end{figure}

\begin{figure}[t]
\includegraphics[width=0.95\columnwidth]{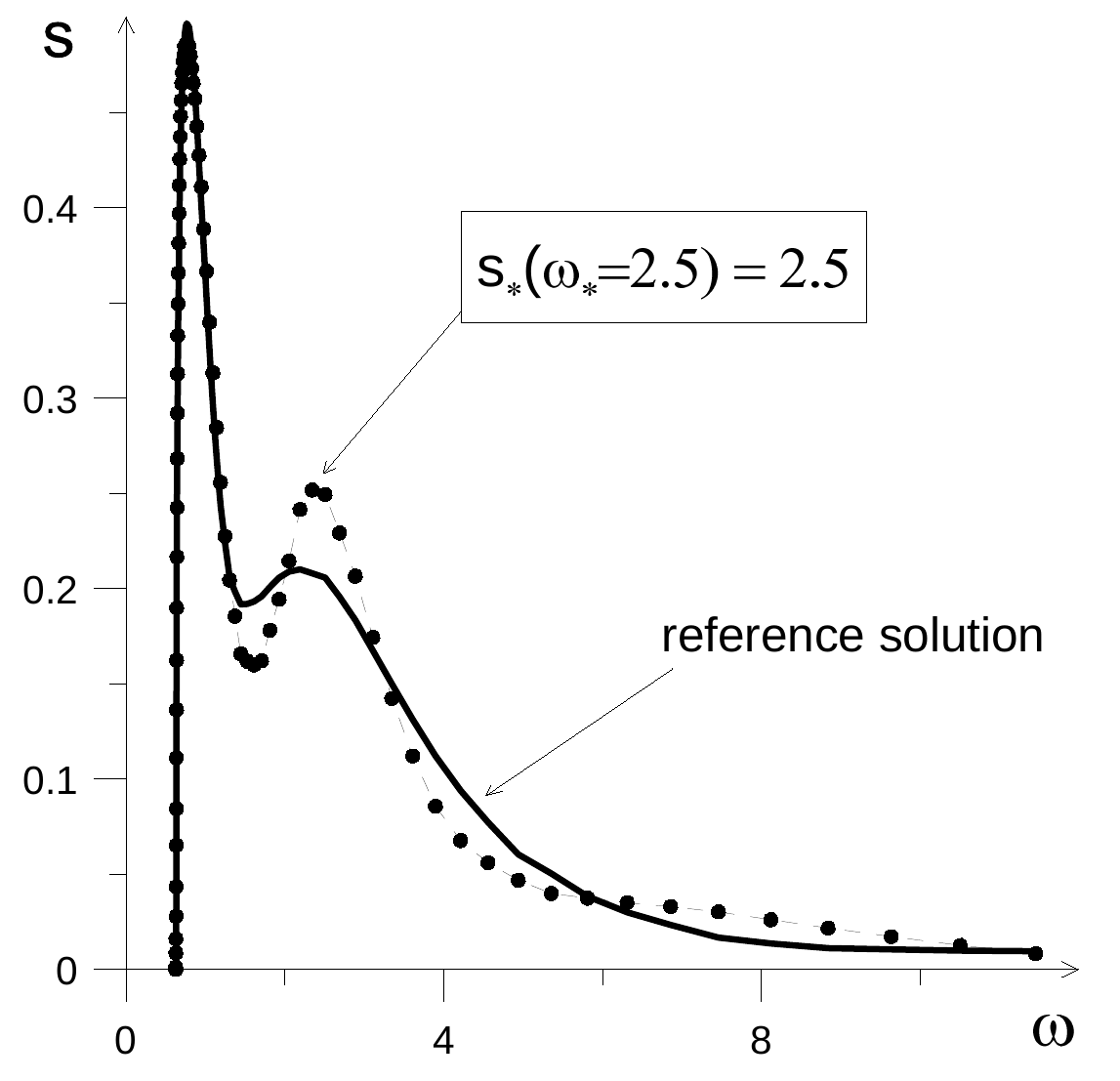}
\vspace*{-0.2cm}
\caption{\label{fig:3} Example of error analysis for the case B. Shown with a solid line is the reference (the smoothest) solution $s(\omega)$,  previously presented in Fig.~\ref{fig:2}.
The solution $s_*(\omega)$ is obtained by pulling the solution up---without compromising the deviation of $g_s(\tau)$ from $s(\omega)$---at the point $\omega=2.5$ corresponding to the
maximum of the second peak. The solution $s_*(\omega)$ corresponds to the threshold of appearance of an extraneous feature, a shoulder at $\omega \approx 6.5$ (that will develop into a bump if we keep pulling up).
In this sense,   $s_*(\omega)$ characterizes maximal potential deviation of the reference solution from the exact one in the vicinity of the second peak, cf. Fig.~\ref{fig:2}.
}
\end{figure}

\begin{figure}[t]
\includegraphics[width=0.95\columnwidth]{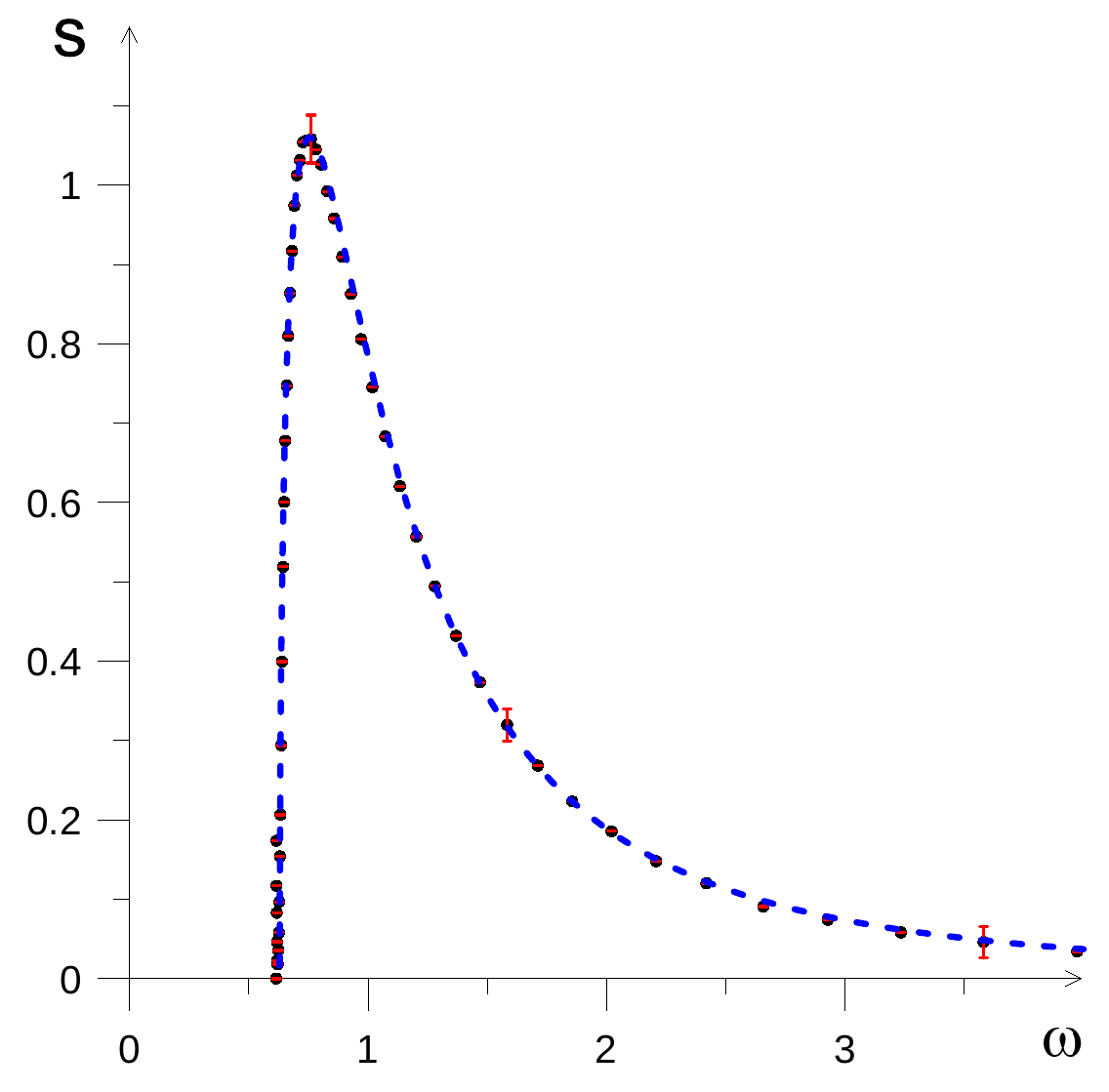}
\vspace*{-0.2cm}
\caption{\label{fig:4}  Case A with three characteristic error bars (cf. Fig.~\ref{fig:1}).
}
\end{figure}

\begin{figure}[t]
\includegraphics[width=0.95\columnwidth]{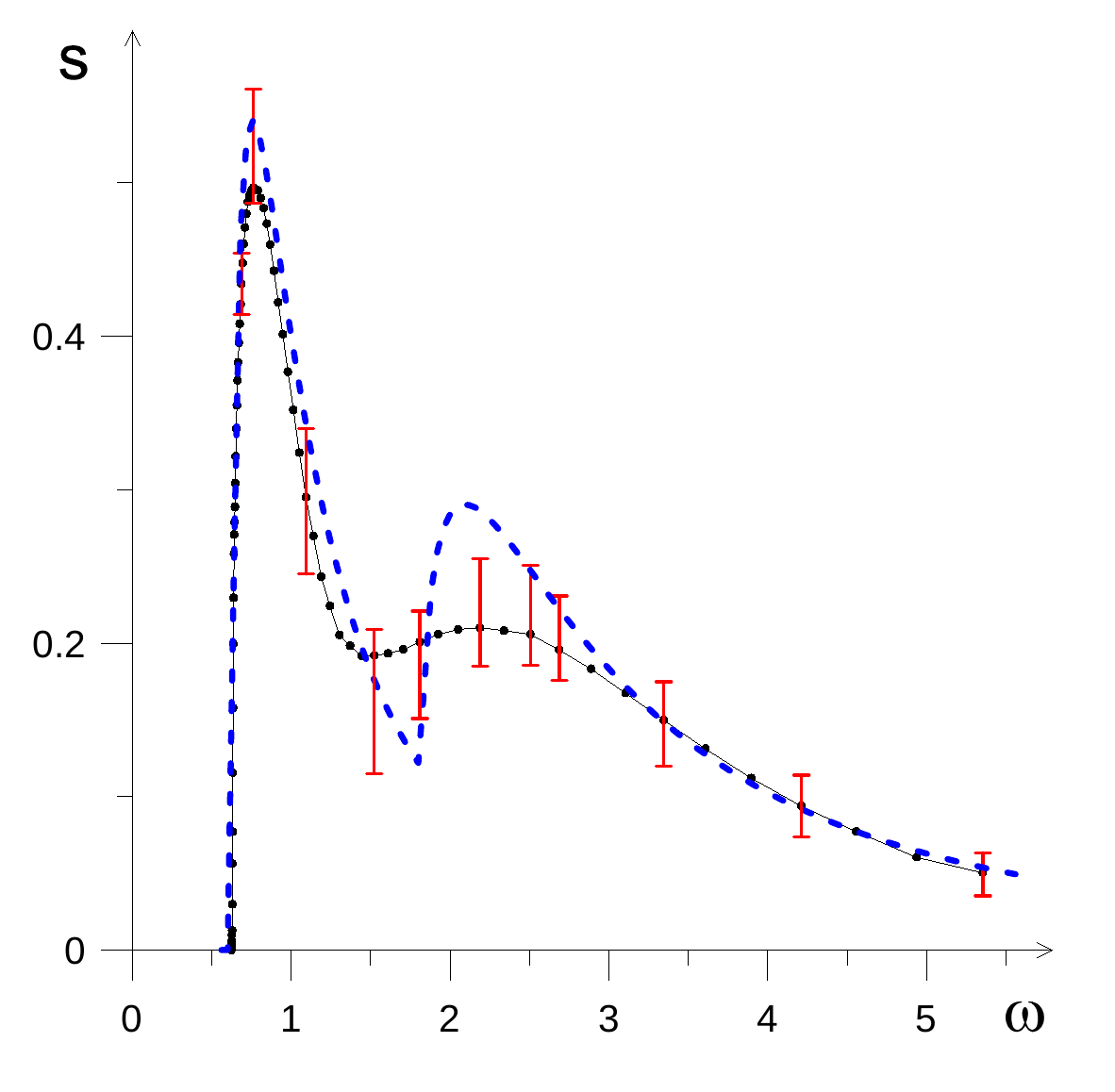}
\vspace*{-0.2cm}
\caption{\label{fig:5}  Case B with the error bars (cf. Figs.~\ref{fig:2} and \ref{fig:3} ). The error bars are essentially asymmetric, reflecting the tendency
of the reference solution  to broaden sharp features.
}
\end{figure}

{\it Illustrative examples.} Leaving technical description of the MCC numeric protocol to the second half of the paper, here we consider two illustrative examples.
Figures \ref{fig:1} and \ref{fig:2} show the results (cases A and B, respectively) of applying MCC to determine spectral functions
for $g(\tau)$ specified numerically with known error bars.
Corresponding functions $s(\omega)$ are shown along with their exact counterparts $s_e (\omega)$. The functions
\begin{equation}
g_s(\tau)\, =\, \int_0^{\infty} {\rm e}^{-\omega \tau}s(\omega) \, d\omega
\label{g_tau_MCC}
\end{equation}
coincide with $g(\tau)$ within the error bars on the latter, while the functions $s(\omega)$ are quite smooth.
Now we have to characterize possible deviations of $s(\omega)$ from $s_e(\omega)$---pretending that the latter
is unknown---making sure that  what we get is consistent with
the deviations we see in the two figures. In Fig.~\ref{fig:3} we show how we quantify the uncertainty of the sharpness of the second peak.
The error bars for both cases are presented in Figs.~\ref{fig:4}  and \ref{fig:5}. The asymmetry of the error bars  reflects the  tendency
of the reference solution  to broaden sharp features. Apart from the error bar asymmetry,  {\it correlations} between the errors are very informative,
as is clear from Fig.~\ref{fig:3}.


{\it Objective function.}
Numerically, the left hand side of Eq.~(\ref{g_tau}) is known on a discrete set of
$\tau$-points $\tau = (\tau_1,\tau_2, \dots , \tau_N)$ and $g(\tau_i)$
values have statistical error bars $\sigma_i$.
The solution for the spectral function will be defined on a dense enough set of frequency points
$\omega =  (\omega_1,\omega_2, \dots , \omega_N)$ so that the integral in Eq.~(\ref{g_tau})
is transformed into the finite-sum expression defining a set of $g_s(\tau_i)$ values
\begin{equation}
g_s(\tau_i) = \sum_{j=1}^{M} \, s_j \, e^{-\omega_j \tau_i } \;.
\label{gstau2}
\end{equation}

The objective function to be minimized in the process of searching for the
smooth solution $s(\omega_j)$ involves several terms $O=\sum_k O_k$. In what follows we
describe the minimal number of objectives which allow one to achieve the final goal
(all results presented in this Letter were based on them).
The first and most important term is the standard $\chi^2$ measure
which penalizes differences $g(\tau_i)-g_s(\tau_i)$ outside of the computed
error bars, $\sigma_i$. For simplicity, we write this measure for the case of uncorrelated statistical errors
\begin{equation}
O_1=N\chi^2 = \sum_{i=1}^{N} \left[ \frac{g(\tau_i)-g_s(\tau_i)}{\sigma_i} \right]^2  .
\label{chi2}
\end{equation}
The major goal is to have this objective of the order of unity.
Penalty for first derivatives is preventing the development
of the saw-instability, i.e. fast changing solutions are disfavored
\begin{equation}
O_2= \sum_{j=1}^{M-1}\,  D_j^2 \, d_j^2 \; .
\label{O2}
\end{equation}
To simplify notations, we introduced $d_j = |s_{j+1}-s_j|$.
Since the minimization of quadratic form does not guarantee that the solution
is non-negative, we also need a penalty which suppresses the amplitudes of the solution
\begin{equation}
O_3= \sum_{j=1}^{M}\,  A_j^2 \, s_j^2 \;.
\label{O3}
\end{equation}
Finally, we introduce penalty for the solution to deviate from some ``target"
form $\bar{s}_j$ for $j=(2,\dots ,M-1)$:
\begin{equation}
O_4= \sum_{j=2}^{M-1} \, T_j^2 \, (s_j-\bar{s}_j)^2 \;.
\label{O4}
\end{equation}

There is nothing new in the idea of introducing regularization measures similar to
$O_2$, or $Q_4$ \cite{Tikhonov,Phillips,BSS,MaxEnt1,MaxEnt1} but in the past it was done with
$j$-independent coefficients considered essentially as input parameters (ultimately
optimized to achieve the best, but somewhat compromised, $\chi^2$). In our scheme,
it is absolutely crucial that constraints control every point of the solution and
are adjusted by the feedback loop to be consistent with the properties of the solution itself.
Only in this case do we have a guarantee that in the limit of vanishingly small error
bars on $g(\tau)$ the final solution will always reach the $\chi^2 \sim 1$ limit.

{\it MCC protocol.} Given an objective based on the positive definite quadratic form for $s_j$,
one can relatively easily find the solution minimizing it, $s_j^{\rm (opt)}(O)$,
e.g. by the gradient method.  This solution, however, may have two serious drawbacks:
It may contain negative values of $s_j$ and be far from meeting the crucial requirement
of having $\chi^2 \sim 1$. The following self-consistent iterative protocol is designed
to eliminate both shortcomings. \\

\noindent 1. Let index $k$ denote the number of performed iterations. Start
with $k=0$, some initial solution $s_j^{(0)}$, large penalties for first derivatives
$D_j^{(0)}$, and zero values of $A_j^{(0)}$.
Define $\bar{s_j}^{(0)} = (s_{j+1}^{(0)}+s_{j-1}^{(0)})/2$ for $j=2,\ldots , M-1$.
We find it sufficient to have $T_j^{(k)}=D_j^{(k)}$ for any $k$ (with small modification
in the protocol quantifying error bars, see below), but one is free to design other
rules for these coefficients. \\

\noindent 2. Determine the next iteration solution by minimizing the objective function,
$s_j^{(k+1)}=s_j^{\rm (opt)}(O^{(k)})$;  set $k\to k+1$.\\

\noindent 3. Since $Q_1$ term is the only
reason for having a non-flat result, the solution is now analyzed to adjust the objective
so that penalties compromising the $\chi^2$ measure are reduced, and penalties for
developing the negative solution are increased:
If $d_j^{(k)}$ exceeds $C/D_j^{(k-1)}$ (in practice, we use $C=0.1$) then
\begin{equation}
D_j^{(k)}= C/d_j^{(k)} \;.
\label{adjustDa}
\end{equation}
Otherwise, the penalty is considered to be too conservative and is increased as
$D_j^{(k)} = 2 D_j^{(k)}$. To prevent divergent behavior, determine $D_{\rm min}=min\{ D_j\}$
and restrict allowed values to $rD_{\rm min}$, where $r$ is some large number (typically of order
of $10^3$). After that set $T_j^{(k)}=D_j^{(k)}$.

\noindent 4. If $s_j^{(k)}<0$ introduce large penalty $A_j^{(k)}=A_{\rm max}$ (it can be as large as $10^8$)
to prevent the solution from going negative in the next iteration. If $s_j^{(k)}>0$
decrease the penalty, $A_j^{(k)}=A_j^{(k-1)}/10$ (for positive values of the solution
$A$-coefficients decay exponentially fast with the number of iterations). \\

\noindent 5. Finally, set  $\bar{s_j}^{(k)} = (s_{j+1}^{(k)}+s_{j-1}^{(k)})/2$ to stabilize second
derivatives of the solution. This determines the new objective function $O^{(k)}$. Proceed to step 2. \\

In essence, the procedure adjusts regularization parameters by feedback from
the solution itself and ultimately has them small enough to admit a
solution of Eq.(\ref{g_tau}) within the error bars, and large enough to have
a smooth solution. If there are $\delta$-functional peaks in the spectral function one
should add them to the solution at some locations and exclude their amplitudes
from the objectives $O_2$ and $O_4$ for obvious reasons. The nature and stability of
the MCC feedback loop is also compatible with interrupting it
from time to time and suggesting modifications to the existing solution which minimize
only the $\chi^2$---the subsequent MCC protocol quickly erases unwanted ripples. We also
find it useful to ``level-off" large point-to-point fluctuations in the regularization
parameters after several feedback loops.

The protocol of examining error bars on $s_j$ is identical to the one described above
except that for some frequency point $\omega_m=\omega_*$ the target parameter $\bar{s}_m$ remains
fixed at some predetermined value $s_*$ and $T_m$ is made large enough to suppress significant
deviations of the solution from $s_*$. As we increase/decrease the value of $s_*$ we ultimately
observe that either the value of $\chi^2$ is increased by a factor of two, or a new feature appears
on the curve. This determines the error margins on the final solution. [Since in the protocol of
determining error bars on $s_j$ the values of $\bar{s}_m$ and $T_m$ are excluded from the feedback
loop the $\chi^2$ measure may increase as we ``pull" the point due to hysteresis effect.]

{\it Discussion.} As is seen from the above-described protocol,  the central idea of MCC is to
iteratively adjust---tighten/relax---regularization constraints, by feedback from 
the solution  $s(\omega)$ that minimizes the objective function for the previous iteration. 
The choice of constraints is rather wide; they can deal with the values of the function $s(\omega)$,
as well as with the values of its first and higher derivatives; within one and the same iteration, or addressing previous iterations as well.
The crucial common feature is the locality of constraints in $\omega$-space, which readily allows one
to judge---see steps 3 and 4 of the protocol---whether constraints are (i) too restrictive,  (ii) too loose, or (iii) perfectly consistent.
Performancewise, it is due to those simple {\it local} measures of consistency of each specific constraint that feedback iterations rapidly fine-tune the objective function to be the
most restrictive---while remaining free of systematic bias---within the error bars of the function $g(\tau)$.

In principle, the set of consistent constraints can be extended to include any knowledge about likely features of the function $s(\omega)$. For example, expecting
a simple monotonic asymptotic behavior in the tail, one may opt to severely penalize wrong signs of (higher) derivatives, similarly to penalizing negative values of $s(\omega)$.

The MCC protocol can be applied to any problem of restoring a function from a representative set of its integrals with an arbitrary kernel. An immediate example is the numeric
function $g(\tau)$ itself, when it comes from a Monte Carlo simulation in 
terms of integrals of $g(\tau)$ over a number of intervals (``bins"). 
An accurate value of $g(\tau)$ for any $\tau$ can then be extracted by the MCC---either directly, or as a by-product of restoring $s(\omega)$ from the bin integrals.

In absorption spectroscopy, MCC can be used for unbiased 
restoring unknown density distribution from binned absorption images, 
or just to produce a smooth---still unbiased (!)---image from data integrated over bins.

This work was supported by the National Science Foundation under grant PHY-1005543,
and by a grant from the Army Research Office with funding from DARPA.

\end{document}